\documentclass[hidelinks,onefignum,onetabnum]{siamart251216}

\usepackage[T1]{fontenc}
\usepackage[utf8]{inputenc}
\usepackage{microtype}

\usepackage{cite}

\usepackage{amssymb}

\usepackage{siunitx}
\DeclareSIUnit{\byte}{B}

\usepackage{booktabs}
\usepackage{multirow}

\usepackage{listings}
\definecolor{codekeyword}{rgb}{0.0, 0.4, 0.65}
\definecolor{codecomment}{rgb}{0.4, 0.45, 0.5}
\definecolor{codestring}{rgb}{0.2, 0.5, 0.25}
\definecolor{codenumber}{rgb}{0.75, 0.75, 0.75}
\definecolor{codebase}{rgb}{0.15, 0.15, 0.15}
\usepackage{algpseudocode}

\usepackage{placeins}

\usepackage{tikz}
\usetikzlibrary{shapes.multipart, arrows.meta, positioning, calc, fit}
\definecolor{lyrA}{rgb}{0.20, 0.50, 0.25}
\definecolor{lyrB}{rgb}{0.02, 0.47, 0.45}
\definecolor{lyrC}{rgb}{0.00, 0.40, 0.65}
\definecolor{lyrD}{rgb}{0.30, 0.28, 0.58}

\newcommand{\authmark}[1]{\textsuperscript{\ensuremath{#1}}}

\headers{Gradient Reconstruction in LBM}%
        {A.~Kummerl\"ander, F.~Bukreev, and M.~J.~Krause}

\title{Gradient Reconstruction in Lattice Boltzmann Methods for
  Systems of Conservation Laws}

\author{%
  Adrian Kummerl\"ander\authmark{*,\dagger}%
  \and Fedor Bukreev\authmark{*,\ddagger}%
  \and Mathias J. Krause\authmark{*,\dagger,\ddagger}%
}
\makeatletter
\g@addto@macro\@thanks{%
  \footnotetext[1]{All authors are with the Lattice Boltzmann Research Group
    (LBRG), Karlsruhe Institute of Technology (KIT), 76131 Karlsruhe, Germany
    (\email{kummerlaender@kit.edu}, corresponding author).}%
  \footnotetext[2]{Institute for Applied and Numerical Mathematics (IANM), KIT,
    Englerstra\ss{}e 2.}%
  \footnotetext[3]{Institute of Mechanical Process Engineering and Mechanics
    (MVM), KIT, Stra\ss{}e am Forum 8.}%
}
\makeatother

\ifpdf
\hypersetup{
  pdftitle={Gradient Reconstruction in Lattice Boltzmann Methods for Systems of Conservation Laws},
  pdfauthor={A. Kummerlaender, F. Bukreev, and M. J. Krause}
}
\fi

\begin{document}

\maketitle

\begin{abstract}
The automatic derivation of~\cite{pde2lbm} turns a declared system of conservation laws into a lattice Boltzmann scheme, giving each conserved physical quantity a set of $q$ populations whose linear equilibrium embeds the physical flux in their first moment.
When the flux depends on gradients of the conserved state, those gradients are supplied by tracking them as additional transported fields.
Since lattice Boltzmann is commonly memory-bound, these additional degrees of freedom reduce the achievable throughput.
To reclaim it, we reconstruct the gradients from the moment structure of the equilibrium instead of transporting them.
To leading order, the first moment of a conserved quantity's non-equilibrium part carries its gradient.
Because the reconstructed flux enters its own equilibrium reference, that moment is the image of the gradient under a linear operator built from the diffusive-flux Jacobian.
The reconstruction is that operator's algebraic inverse, generic across gradient-form constitutive closures and generated automatically from each declared flux.
The resulting scheme carries the conserved quantities alone, reconstructing the required gradients from the populations and forming the fluxes locally.
It converges at second order in double precision across advection--diffusion--reaction, Allen--Cahn, Navier--Stokes, resistive magnetohydrodynamics and homogenized compressible Navier--Stokes--Fourier systems, matching the accuracy of gradient tracking at equal resolution.
On an NVIDIA RTX A5000 it is up to $3.7$ times faster in single precision, the memory-bound kernels reaching up to $97\%$ of peak memory bandwidth.
\end{abstract}

\begin{keywords}
  Lattice Boltzmann methods, gradient reconstruction, non-equilibrium moment, method of manufactured solutions, high-performance computing
\end{keywords}

\begin{MSCcodes}
  76M28, 65M12, 68W30, 65Y05, 35L65
\end{MSCcodes}

\section{Introduction}
The derivation in~\cite{pde2lbm} produces a lattice Boltzmann scheme from a declared system of conservation laws, targeting OpenLB~\cite{Krause2021a, Kummerlaender2023}.
It assigns an independent set of populations to each conserved field and embeds the physical flux in the first moment of a linear equilibrium, the vector-kinetic and relaxation route to systems of conservation laws~\cite{JinXin1995, ChenLevermoreLiu1994, Natalini1998, Bouchut1999, BouchutGuarguagliniNatalini2000, AregbaNatalini2000, LattanzioNatalini2002, AregbaNataliniTang2004, Graille2014, Dubois2014}.
Wherever a constitutive closure is a first gradient of the conserved state, as for the Newtonian stress and the Fourier flux, the scheme needs that gradient.
It has so far been provided by transport, in one of two forms~\cite{pde2lbm, Bukreev2026cnsf, Bukreev2026}:
\emph{Gradient tracking} carries the gradient itself.
The \emph{relaxed-flux} form carries the diffusive flux itself: the viscous stress and the heat flux.
Both append an auxiliary field, advanced by relaxation alongside the conserved ones, and differ only in what that field represents.
Since lattice Boltzmann is commonly bound by memory bandwidth rather than floating-point performance~\cite{Krueger2016, Tolke2010, Januszewski2014, Godenschwager2013, Bauer2021walberla, Bauer2021lbmpy}, the additional degree of freedom carried by the auxiliary field directly reduces the achievable throughput.
Yet the field is redundant: while carried as an independent degree of freedom, it holds no information the conserved state does not already determine, because the closure ties the flux to a gradient of that state.

The present work recognizes these independent degrees of freedom as unnecessary and removes them.
This route was noted in~\cite{pde2lbm} but not pursued.
To leading order, the first moment of a conserved field's non-equilibrium part carries that field's gradient, by the Chapman--Enskog relation that recovers strain rate and heat flux from the non-equilibrium populations~\cite{Latt2006, Malaspinas2015, Coreixas2017, Krafczyk2003, JunkKlarLuo2005, ShanYuanChen2006, Krueger2016}.
As such, the gradients a closure needs can be reconstructed inside the collision from the conserved fields alone.
The reconstruction is a single declaration, generated from the declared flux, trading transport for local computation.

Unlike the transport forms, where the gradient is read from its own transported field, the reconstructed gradient feeds back into its own measurement.
The diffusive flux built from it is injected into the equilibrium first moment, the reference against which the next non-equilibrium moment is taken.
The moment therefore returns not the gradient but its image under a linear operator built from the diffusive-flux Jacobian.
Taken directly as the gradient, it carries an error that does not vanish under refinement.

The operator is linear, so the reconstruction is its algebraic inverse.
Its coefficient is already assembled symbolically by the compiler of~\cite{pde2lbm} with the SymPy computer algebra system~\cite{Meurer2017}, so the inverse can be generated in closed form from each declared flux.
We verify the scheme by manufactured solutions across the five gradient-tracking systems of~\cite{pde2lbm} and by a throughput study on an NVIDIA RTX A5000 GPU.

The remainder of this paper is organized as follows:
Section~\ref{sec:methodology} presents the method: Section~\ref{sec:chapman_enskog} states the gradient reconstruction, Section~\ref{sec:feedback} derives the reconstruction operator and Section~\ref{sec:generation} describes its generation by the compiler.
Section~\ref{sec:convergence} verifies the scheme by the method of manufactured solutions, and Section~\ref{sec:performance} benchmarks its throughput against the gradient-tracking scheme and hardware roofline.

\section{Methodology}\label{sec:methodology}

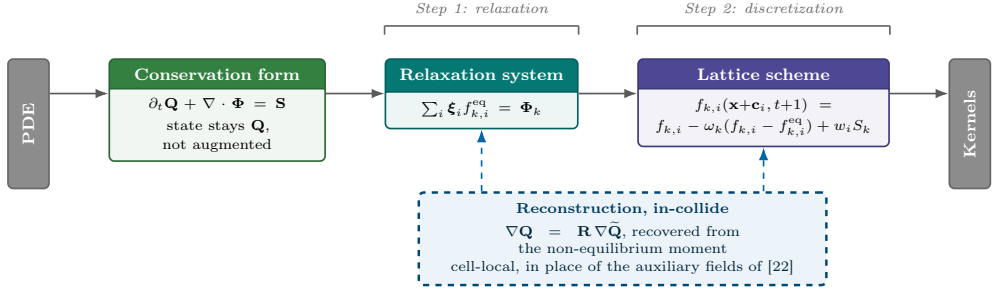
\begin{figure}[t]
\centering
\resizebox{\textwidth}{!}{%
\begin{tikzpicture}[
  font=\footnotesize,
  stage/.style={rectangle split, rectangle split parts=2, rounded corners=2.5pt,
                rectangle split part fill={#1,#1!7},
                draw=#1!80!black, thick, align=center, text width=30mm, inner sep=4pt},
  vbox/.style={rectangle, rounded corners=2.5pt, fill=#1, draw=black!60,
               thick, inner sep=3pt, minimum height=22mm},
  arr/.style={-{Latex[length=2.2mm]}, thick, black!65},
  brk/.style={black!60, thick},
]
\node[vbox=black!45, minimum width=7mm] (b0) {\rotatebox{90}{\textcolor{white}{\textbf{PDE}}}};
\node[stage=lyrA, anchor=north west, text width=34mm] (b1) at ([xshift=10mm]b0.north east)
  {\textcolor{white}{\textbf{Conservation form}}
  \nodepart{two}{\scriptsize $\partial_t \mathbf{Q} + \nabla\cdot\boldsymbol{\Phi} = \mathbf{S}$\\[1pt]
    state stays $\mathbf{Q}$, not augmented}};
\node[stage=lyrB, anchor=north west] (b2) at ([xshift=10mm]b1.north east)
  {\textcolor{white}{\textbf{Relaxation system}}
  \nodepart{two}{\scriptsize $\textstyle\sum_i \boldsymbol{\xi}_i f^{\mathrm{eq}}_{k,i} = \boldsymbol{\Phi}_k$}};
\node[stage=lyrD, anchor=north west, text width=40mm] (b3) at ([xshift=10mm]b2.north east)
  {\textcolor{white}{\textbf{Lattice scheme}}
  \nodepart{two}{\scriptsize $f_{k,i}(\mathbf{x}{+}\mathbf{c}_i, t{+}1) =$\\[1pt]
    $f_{k,i} - \omega_k(f_{k,i}-f^{\mathrm{eq}}_{k,i}) + w_i S_k$}};
\node[vbox=black!45, anchor=north west, minimum width=7mm] (b4) at ([xshift=10mm]b3.north east)
  {\rotatebox{90}{\textcolor{white}{\textbf{Kernels}}}};
\draw[arr] ([yshift=-6mm]b0.north east) -- ([yshift=-6mm]b1.north west);
\draw[arr] ([yshift=-6mm]b1.north east) -- ([yshift=-6mm]b2.north west);
\draw[arr] ([yshift=-6mm]b2.north east) -- ([yshift=-6mm]b3.north west);
\draw[arr] ([yshift=-6mm]b3.north east) -- ([yshift=-6mm]b4.north west);
\draw[brk] ([yshift=5.5mm]b0.north west -| b2.west) -- ([yshift=5.5mm]b0.north west -| b2.east)
  node[midway, above=1pt, font=\scriptsize\itshape, text=black!60]{Step 1: relaxation};
\draw[brk] ([yshift=5.5mm]b0.north west -| b2.west) -- ([yshift=4mm]b0.north west -| b2.west);
\draw[brk] ([yshift=5.5mm]b0.north west -| b2.east) -- ([yshift=4mm]b0.north west -| b2.east);
\draw[brk] ([yshift=5.5mm]b0.north west -| b3.west) -- ([yshift=5.5mm]b0.north west -| b3.east)
  node[midway, above=1pt, font=\scriptsize\itshape, text=black!60]{Step 2: discretization};
\draw[brk] ([yshift=5.5mm]b0.north west -| b3.west) -- ([yshift=4mm]b0.north west -| b3.west);
\draw[brk] ([yshift=5.5mm]b0.north west -| b3.east) -- ([yshift=4mm]b0.north west -| b3.east);
\node[draw=lyrC!70!black, dashed, very thick, rounded corners=2.5pt,
      align=center, inner sep=5pt, text width=66mm, fill=lyrC!7, font=\scriptsize,
      text=lyrC!35!black, below=9mm of $(b2.south)!0.5!(b3.south)$] (rec)
  {\textbf{\textcolor{lyrC!60!black}{Reconstruction, in-collide}}\\[2pt]
   $\nabla\mathbf{Q}=\mathbf{R}\,\nabla\widetilde{\mathbf{Q}}$, recovered from the non-equilibrium moment\\[1pt]
   cell-local, in place of the auxiliary fields of~\cite{pde2lbm}};
\draw[arr, dashed, lyrC] (rec.north -| b2.south) -- (b2.south);
\draw[arr, dashed, lyrC] (rec.north -| b3.south) -- (b3.south);
\end{tikzpicture}%
}
\caption{The compiler pipeline with the gradient reconstruction in place of the gradient-tracking
cascade of~\cite{pde2lbm}. A declared conservation law keeps its conserved state $\mathbf{Q}$. The
reconstruction operator $\mathbf{R}$ recovers the gradient $\nabla\mathbf{Q}$ inside the collision
(Sections~\ref{sec:chapman_enskog}--\ref{sec:feedback}), so the relaxation and lattice stages act on
$\mathbf{Q}$ alone rather than on an augmented state $\mathbf{Q}^{+} = (\mathbf{Q}, \nabla\mathbf{Q})$.}
\label{fig:pipeline}
\end{figure}

Figure~\ref{fig:pipeline} shows the overall pipeline: a declared system of conservation laws is compiled to a lattice Boltzmann scheme, with the constitutive gradients reconstructed inside the collision rather than tracked as auxiliary fields.

\subsection{Gradient Reconstruction}\label{sec:chapman_enskog}

The method of~\cite{pde2lbm} takes a system of conservation laws as its entry point,
\begin{equation}\label{eq:system}
\partial_t \mathbf{Q} + \nabla\cdot\boldsymbol{\Phi} = \mathbf{S},
\end{equation}
with $\mathbf{Q}$ the conserved state, $\boldsymbol{\Phi}$ the physical flux and $\mathbf{S}$ a local source.
Each conserved field $k$ with moment $Q_k$ and flux $\boldsymbol{\Phi}_k$ is assigned an independent set of populations $f_{k,i}$ on a discrete-velocity stencil.
The flux is embedded in the first moment of a linear equilibrium, the field is recovered from the zeroth moment, and the scheme iterates by a BGK collide-and-stream update with the source projected onto the lattice weights.

The equilibrium $f_{k,i}^{\mathrm{eq}}$ of~\cite{pde2lbm} is linear in the state and reproduces the conserved moment, the physical flux and an isotropic second moment,
\begin{equation}\label{eq:eqmoments}
  \sum_i f_{k,i}^{\mathrm{eq}} = Q_k, \qquad
  \sum_i f_{k,i}^{\mathrm{eq}}\, \boldsymbol{\xi}_i = \boldsymbol{\Phi}_k, \qquad
  \sum_i f_{k,i}^{\mathrm{eq}}\, \boldsymbol{\xi}_i \otimes \boldsymbol{\xi}_i = c_s^2 Q_k\, \mathbf{I},
\end{equation}
with $\boldsymbol{\xi}_i = \mathbf{c}_i$ the lattice velocities and $c_s$ the lattice speed of sound.
The Chapman--Enskog expansion $f_{k,i} = f_{k,i}^{\mathrm{eq}} + \Delta t\, f_{k,i}^{(1)} + \mathcal{O}(\Delta t^2)$, $\sum_i f_{k,i}^{(1)} = 0$ relates the non-equilibrium part, with $\tau_{\mathrm{LB},k}$ the relaxation time of field $k$, to the derivative of the equilibrium along the discrete velocity $\boldsymbol{\xi}_i$,
\begin{equation}\label{eq:f1}
  f_{k,i}^{(1)} = -\tau_{\mathrm{LB},k} \left( \partial_t + \boldsymbol{\xi}_i \cdot \nabla \right)
  f_{k,i}^{\mathrm{eq}} + \mathcal{O}(\Delta t).
\end{equation}
Its first moment, evaluated with the equilibrium moments~\eqref{eq:eqmoments}, is the non-equilibrium deviation flux
\begin{equation}\label{eq:pi1}
\begin{split}
    \boldsymbol{\Pi}_k^{(1)}
    &= \sum_i f_{k,i}^{(1)}\, \boldsymbol{\xi}_i \\
    &= -\tau_{\mathrm{LB},k} \left( \partial_t \boldsymbol{\Phi}_k + \nabla\cdot\!\big(c_s^2 Q_k \mathbf{I}\big) \right) \\
    &= -\tau_{\mathrm{LB},k} \left( \partial_t \boldsymbol{\Phi}_k + c_s^2\, \nabla Q_k \right).
\end{split}
\end{equation}
In~\cite{pde2lbm} this relation supplies the relaxation diffusion.
The flux time derivative is closed by the chain rule
\begin{equation}\label{eq:chainclose}
  \partial_t \boldsymbol{\Phi}_k = \mathbf{A}_k\, \partial_t \mathbf{Q}
  = \mathbf{A}_k\,(\mathbf{S} - \nabla\cdot\boldsymbol{\Phi}),
  \qquad \mathbf{A}_k = \frac{\partial \boldsymbol{\Phi}_k}{\partial \mathbf{Q}},
\end{equation}
where the second equality uses $\partial_t \mathbf{Q} = \mathbf{S} - \nabla\cdot\boldsymbol{\Phi}$ from~\eqref{eq:system}.
This closure is algebraic and transports no field.
A constitutive gradient is different: where the flux depends on $\nabla V$ for a primitive $V$, the augmented construction appends an auxiliary field $\mathbf{G} \approx \nabla V$ and relaxes it alongside the conserved state.
We instead set $\partial_t \boldsymbol{\Phi}_k = 0$.
This ties the gradient to the non-equilibrium moment and leaves no independent degree of freedom to transport.

Equation~\eqref{eq:pi1} then reads $\boldsymbol{\Pi}_k^{(1)} = -c_s^2\, \tau_{\mathrm{LB},k}\, \nabla Q_k$, so the gradient would follow as $\nabla Q_k = -\boldsymbol{\Pi}_k^{(1)} /(c_s^2 \tau_{\mathrm{LB},k})$.
This non-equilibrium moment $\boldsymbol{\Pi}_k^{(1)} = \sum_i f_{k,i}\,\boldsymbol{\xi}_i - \boldsymbol{\Phi}_k$, in lattice units where $\Delta t = 1$, subtracts the full equilibrium flux, but $\boldsymbol{\Phi}_k$ carries the gradient in its diffusive part and cannot be formed in-collide.
Subtracting only its advective part $\boldsymbol{\Phi}_k^{\mathrm{adv}} = \boldsymbol{\Phi}_k|_{\nabla \mathbf{Q} = 0}$, a function of the conserved state $Q_k = \sum_i f_{k,i}$ alone, gives the direct reconstruction
\begin{equation}\label{eq:gradrecon}
    \nabla \widetilde{Q}_k = -\frac{1}{c_s^2\, \tau_{\mathrm{LB},k}} \left( \sum_i f_{k,i}\,
    \boldsymbol{\xi}_i - \boldsymbol{\Phi}_k^{\mathrm{adv}} \right).
\end{equation}
Both terms are known during the collision, so $\nabla \widetilde{Q}_k$ is available in-collide for every conserved field.
The diffusive flux it still leaves in the moment is what Section~\ref{sec:feedback} accounts for.

Recovering the gradient from the non-equilibrium moment while neglecting the flux time derivative is the local strain-rate and heat-flux evaluation of regularized and non-equilibrium lattice Boltzmann methods~\cite{Latt2006, Malaspinas2015, Coreixas2017}.
The neglected term is higher order under the diffusive and acoustic scalings, and the second-order convergence of Section~\ref{sec:convergence} confirms it across the present systems.

The reconstructed gradients are the inputs the constitutive closures require, evaluated in Section~\ref{sec:convergence} across the declared systems, most extensively for the resistive \emph{Magnetohydrodynamics} (MHD) system, which couples a Newtonian stress, a Fourier flux and a resistive term through the reconstructed velocity, temperature and magnetic-field gradients.

\subsection{The Reconstruction Operator}\label{sec:feedback}

The diffusive flux left in the direct reconstruction~\eqref{eq:gradrecon} is the remainder of the physical flux after its advective part, $\boldsymbol{\Phi}^{\mathrm{diff}} = \boldsymbol{\Phi} - \boldsymbol{\Phi}^{\mathrm{adv}}$.
Because the constitutive closures are first gradients of the state, it is linear in the gradient,
\begin{equation}\label{eq:coeff}
  \boldsymbol{\Phi}^{\mathrm{diff}} = \mathbf{M}\,\nabla\mathbf{Q}, \qquad
  \mathbf{M} = \frac{\partial \boldsymbol{\Phi}^{\mathrm{diff}}}{\partial(\nabla\mathbf{Q})},
\end{equation}
with $\mathbf{M}$ the diffusive-flux Jacobian.
The equilibrium first moment~\eqref{eq:eqmoments} carries $\mathbf{M}\,\nabla\mathbf{Q}$, so with $\omega_k = 1/\tau_{\mathrm{LB},k}$ the relaxation frequency of field $k$, the direct reconstruction is
\begin{equation}\label{eq:feedbackmap}
  \nabla\widetilde{\mathbf{Q}}
  = \left( \mathbf{I} - c_s^{-2}\, \mathrm{diag}(\omega_k)\, \mathbf{M} \right) \nabla\mathbf{Q}
\end{equation}
rather than $\nabla\mathbf{Q}$ itself.
The true gradient is the inverse,
\begin{equation}\label{eq:feedback}
    \nabla \mathbf{Q} = \mathbf{R}\, \nabla\widetilde{\mathbf{Q}}, \qquad
    \mathbf{R} = \left( \mathbf{I} - c_s^{-2}\, \mathrm{diag}(\omega_k)\, \mathbf{M} \right)^{-1},
\end{equation}
the reconstruction operator $\mathbf{R}$, assembled from the diffusive-flux Jacobian $\mathbf{M}$ and the field relaxation frequencies.
The form $\mathbf{R}$ takes follows from the closure, and separates the systems into two cases.
For a scalar diffusive closure, flux $-D\,\nabla c$, the Jacobian~\eqref{eq:coeff} is the constant $\mathbf{M} = -D\,\mathbf{I}$, and $\mathbf{R}$ reduces to a single scaling factor
\begin{equation}\label{eq:scalarcorr}
  \mathbf{R} = \frac{1}{1 + \beta}\,\mathbf{I}, \qquad
  \beta = c_s^{-2}\,\omega\,D_{\mathrm{lat}}, \qquad
  D_{\mathrm{lat}} = D\,\frac{\Delta t}{\Delta x^2},
\end{equation}
with $D_{\mathrm{lat}}$ the diffusivity in lattice units.
The coefficient is constant, so this $\mathbf{R}$ is formed once at compile time.
The second, state-dependent case is a Newtonian stress $\boldsymbol{\tau} = \mu(\nabla\mathbf{u} + \nabla\mathbf{u}^{\mathsf{T}} - \tfrac{2}{3}(\nabla\!\cdot\mathbf{u})\,\mathbf{I})$.
By the chain rule $\nabla\mathbf{u} = (\nabla(\rho\mathbf{u}) - \mathbf{u}\otimes\nabla\rho)/\rho$, its Jacobian is $\mathbf{M} = -(\mu/\rho)\,\mathsf{S}$, with $\mathsf{S}$ the symmetric-deviatoric map $\mathsf{S}\,\nabla\mathbf{u} = \nabla\mathbf{u} + \nabla\mathbf{u}^{\mathsf{T}} - \tfrac{2}{3}(\nabla\!\cdot\mathbf{u})\,\mathbf{I}$, which acts as the factor two on each shear pair.
The shear block therefore keeps the scalar form $1/(1+\beta)$, but its coefficient is now set by the local density through $\nu = \mu/\rho$, and by the local temperature where $\mu = \mu(T)$.
There $\mathbf{R}$ is not a single constant but is evaluated per cell inside the collision (Section~\ref{sec:generation}).
The advection--diffusion--reaction and Allen--Cahn systems fall in the first case, the Navier--Stokes, magnetohydrodynamics and \emph{Homogenized Compressible Navier--Stokes--Fourier} (HCNSF) systems in the second.

In both cases the correction is significant.
At the diffusive working points, with $c_s^2 = 1/3$ and relaxation near the stability limit $\omega \to 2$, $\beta \approx 2.9$ for advection--diffusion--reaction ($D_{\mathrm{lat}} = 0.48$), $3.8$ for Allen--Cahn, and $2.9$ for the Navier--Stokes shear (there $\beta = 2\nu_{\mathrm{lat}}\,\omega/c_s^2$ with the lattice kinematic viscosity $\nu_{\mathrm{lat}} = \nu\,\Delta t/\Delta x^2$, $\nu = \mu/\rho$, quoted at the reference density $\rho_0$ that eliminates the mass dimension), a scaling factor $1/(1+\beta) \approx 0.26$ ($0.21$ for Allen--Cahn).
Under acoustic scaling $\nu_{\mathrm{lat}} = \nu\,f N$ grows with the resolution $N$ ($f = \Delta t/\Delta x$ fixed), so $\beta \approx 1.5$ for magnetohydrodynamics and $2.0$ for the HCNSF system at the finest grids, factors $0.40$ and $0.33$.
Because $\mathbf{M}$ does not decay in lattice units under refinement, this correction never vanishes: read directly as the gradient, the moment overshoots it several-fold and does not converge to the declared solution.

\subsection{Compilation}\label{sec:generation}

In the compiler the reconstruction is generated for any declared flux rather than hand-written per system.
The call \lstinline|reconstruct_gradients| takes the quantities $V_1, \ldots, V_n$ whose gradients the constitutive closures need and recovers each from the non-equilibrium moment rather than from an auxiliary field, dropping that field.
Each argument must reduce to a function of the conserved state, so that its gradient can be reconstructed from the conserved populations, and a quantity that does not is rejected at compile time.
The pass takes the place of the gradient-tracking cascade of~\cite{pde2lbm}: the conserved state is not augmented, and the required gradients are recovered locally instead (Algorithm~\ref{alg:reconstruct}, Figure~\ref{fig:pipeline}).

\begin{algorithm}[tbp]
\caption{Gradient reconstruction compiler pass.}
\label{alg:reconstruct}
\begin{algorithmic}[1]
\Require flux $\boldsymbol{\Phi}(\mathbf{Q}, \nabla\mathbf{Q})$, discrete velocities $\mathbf{c}_i$, relaxation freq. $\omega_k = 1/\tau_{\mathrm{LB},k}$
\Ensure in-collide reconstruction of $\nabla\mathbf{Q}$, no auxiliary state carried
\Statex \textbf{Split} (symbolic) \hfill \textit{$\boldsymbol{\Phi} \to (\boldsymbol{\Phi}^{\mathrm{adv}}, \boldsymbol{\Phi}^{\mathrm{diff}})$}
\State $\boldsymbol{\Phi}^{\mathrm{adv}} \gets \boldsymbol{\Phi}|_{\nabla\mathbf{Q}=0}$, a function of $\mathbf{Q}$ alone
\State $\boldsymbol{\Phi}^{\mathrm{diff}} \gets \boldsymbol{\Phi} - \boldsymbol{\Phi}^{\mathrm{adv}}$
\Statex \textbf{Form} (symbolic) \hfill \textit{diffusive-flux Jacobian}
\State $\mathbf{M} \gets \partial\boldsymbol{\Phi}^{\mathrm{diff}} / \partial(\nabla\mathbf{Q})$ by CAS differentiation~\cite{Meurer2017}
\Statex \textbf{Assemble} (symbolic) \hfill \textit{reconstruction operator}
\State $\mathbf{R} \gets (\mathbf{I} - c_s^{-2}\,\mathrm{diag}(\omega_k)\,\mathbf{M})^{-1}$ in closed form
\If{$\mathbf{M}$ is constant}
  \State $\mathbf{R}$ is a compile-time constant
\Else
  \State $\mathbf{R}$ is emitted as per-cell block solves
\EndIf
\Statex \textbf{Reconstruct} (per cell, in-collide) \hfill \textit{$f_{k,i} \to \nabla\mathbf{Q}$}
\For{each conserved field $k$}
  \State $\nabla\widetilde{Q}_k \gets -\big(\sum_i f_{k,i}\,\boldsymbol{\xi}_i - \boldsymbol{\Phi}^{\mathrm{adv}}_k\big) / (c_s^2\,\tau_{\mathrm{LB},k})$
\EndFor
\State $\nabla\mathbf{Q} \gets \mathbf{R}\,\nabla\widetilde{\mathbf{Q}}$
\State evaluate the constitutive closures from $\nabla\mathbf{Q}$
\end{algorithmic}
\end{algorithm}

The gradients the closures require are those of the declared primitives $V$, not of the conserved state $\mathbf{Q}$ directly.
The compiler reconstructs the conserved-field gradients $\nabla\mathbf{Q}$ from the non-equilibrium moment (Section~\ref{sec:chapman_enskog}) and forms each requested gradient by the chain rule
\begin{equation}\label{eq:chainrule}
  \nabla V = \frac{\partial V}{\partial \mathbf{Q}}\, \nabla\mathbf{Q},
\end{equation}
with $\partial V/\partial\mathbf{Q}$ the Jacobian of the primitive in the conserved state, as in the Newtonian velocity gradient of Section~\ref{sec:feedback}.
Because the physical flux $\boldsymbol{\Phi}$ is already held symbolically, the compiler forms this Jacobian together with the diffusive-flux Jacobian $\mathbf{M}$ and the reconstruction operator $\mathbf{R}$ of~\eqref{eq:feedback} in closed form in SymPy~\cite{Meurer2017}, the same way for every declared flux with a diffusive part.
Listing~\ref{lst:mhd} declares the resistive magnetohydrodynamics system of Section~\ref{sec:convergence} in this form, its Newtonian stress, Fourier flux and resistive term written directly as first gradients and closed by the single call \lstinline|reconstruct_gradients([u, Temp, B])|.

\begin{lstlisting}[language=Python, caption={DSL formulation of resistive MHD with gradient reconstruction. \lstinline|B| is the Alfv\'en-scaled magnetic field and \lstinline|S_rhou|, \lstinline|S_E|, \lstinline|S_B| are the Powell $\nabla\cdot\mathbf{B}$-cleaning sources~\cite{Powell1999}.}, label={lst:mhd}, float=htbp]
from pde2lbm import *
eqs = ConservationLaws(dim=2)

# conserved fields
rho  = eqs.state("rho",  dim=mass / length**3)
rhou = eqs.vector("rhou", dim=momentum / length**3)
E    = eqs.state("E",    dim=energy / length**3)
B    = eqs.vector("B",   dim=velocity)

# parameters
gamma = eqs.parameter("gamma")
mu    = eqs.parameter("mu",  dim=pressure * time)
K     = eqs.parameter("K",   dim=pressure * time)
eta   = eqs.parameter("eta", dim=length**2 / time)

# primitives
u     = rhou / rho
p_gas = (gamma - 1) * (E - 0.5 * rho * u.dot(u) - 0.5 * B.dot(B))
p_tot = p_gas + 0.5 * B.dot(B)
Temp  = p_gas / rho

# gradient-form closures: Newtonian stress, Fourier flux, resistive term
tau   = mu * (grad(u) + grad(u).T - Rational(2,3) * div(u) * eye(2))
q     = -K * grad(Temp)
E_r   = eta * curl(B)

# conservation laws
eqs.add([
    Eq(dt(rho)  + div(rhou), 0),
    Eq(dt(rhou) + div(outer(rhou,u) + p_tot*eye(2) - outer(B,B) - tau), S_rhou),
    Eq(dt(E) + div((E + p_tot)*u - B*u.dot(B) - tau*u + q + cross(E_r,B)), S_E),
    Eq(dt(B)    + div(outer(B,u) - outer(u,B) - eta*grad(B)), S_B),
])

# recover the closure gradients from the non-equilibrium moment
eqs.reconstruct_gradients([u, Temp, B])
eqs.compile(class_name="ResistiveMHDDynamics")
\end{lstlisting}

Where $\mathbf{M}$ is state-dependent, $\mathbf{R}$ is solved per cell.
This per-cell solve is computationally affordable because $\mathbf{M}$ is a block-sparse matrix: a few small strongly coupled blocks, with the remaining couplings acyclic.
$\mathbf{R}$ is then a few small block solves and scalar divisions by $1 + \beta$, ordered by dependence, rather than a dense inverse, at a cost of order the field count per cell.
In any case, these operations are hidden behind the population loads and stores of the memory-bound kernel, so the reconstruction approaches the memory-bandwidth roofline in single precision (Section~\ref{sec:performance}).

\section{Verification by Manufactured Solutions}\label{sec:convergence}

We verify the reconstruction operator by the method of manufactured solutions on the five gradient-tracking systems of~\cite{pde2lbm}, reusing their exact solutions, lattices and refinement laws so that the errors are directly comparable at equal resolution.
For each system the compiler evaluates the continuous residual of the analytical profile to form the manufactured source $\mathbf{S}_{\mathrm{MMS}} = \partial_t \mathbf{Q}_{\mathrm{exact}} + \nabla\cdot\boldsymbol{\Phi}(\mathbf{Q}_{\mathrm{exact}}) - \mathbf{S}(\mathbf{Q}_{\mathrm{exact}})$, which is injected into the update.
At the final time $t = 1$ we report the relative $L_2$ and $L_\infty$ errors per component and take the \emph{Empirical Order of Convergence} (EOC) to be the negative least-squares slope of $\log_2 \lVert e \rVert$ against $\log_2 N$ over the resolution sequence $N \in \{64, 128, 256, 512\}$, and $N \in \{32, 64, 128, 256\}$ for the three-dimensional HCNSF system.
The only change from the gradient-tracking verification is that the auxiliary gradient is replaced by the reconstruction of Section~\ref{sec:chapman_enskog} with the operator of Section~\ref{sec:feedback}.
Each system is refined under its declared scaling, diffusive for the advection--diffusion--reaction, Allen--Cahn and Navier--Stokes systems and acoustic for the resistive magnetohydrodynamics and the three-dimensional HCNSF systems, matching~\cite{pde2lbm}.
The HCNSF system adds a moving-solid drag source and a porosity-weighted diffusive flux to the compressible equations.
The impact of reconstruction onto the \emph{Compressible Navier--Stokes--Fourier} (CNSF) system we previously~\cite{Bukreev2026cnsf} validated on shock-tube and Taylor--Green vortex flows will be treated separately.
The purely hyperbolic systems of the approach carry no diffusive flux and no gradient closure to reconstruct and are outside the scope of the operator.

\begin{table}[tbp]
\centering
\scriptsize
\setlength{\tabcolsep}{2.5pt}
\renewcommand{\arraystretch}{1.15}
\caption{Manufactured-solution convergence of the gradient reconstruction operator.
Relative $L_\infty$ and $L_2$ errors at the final time $t = 1$ and the finest resolution $N_{\max}$,
with the EOC the negative least-squares log-log slope over a factor-two ladder of four resolutions
ending at the tabulated $N_{\max}$ ($512$, or $256$ for the three-dimensional HCNSF system).
The manufactured solutions, lattices and refinement laws are those of the gradient-tracking
verification~\cite{pde2lbm}. Double and single precision use one relaxation working point per system.}
\label{tab:mms_eoc}
\begin{tabular}{l l l l c c c c c}
\toprule
System & Scaling & Field & Precision & $N_{\max}$ & Rel.\ $L_\infty$ & $L_\infty$ EOC & Rel.\ $L_2$ & $L_2$ EOC \\
\midrule
\multirow{2}{*}{Scalar ADR} & \multirow{2}{*}{Diffusive} & $c$ & Double & 512 & \num{1.55e-04} & $2.00$ & \num{1.03e-04} & $2.00$ \\*
 &  &  & Float & 512 & \num{3.60e-04} & $1.63$ & \num{3.40e-04} & $1.48$ \\
\midrule
\multirow{2}{*}{Allen--Cahn} & \multirow{2}{*}{Diffusive} & $\phi$ & Double & 512 & \num{5.68e-04} & $1.99$ & \num{5.89e-04} & $1.99$ \\*
 &  &  & Float & 512 & \num{1.11e-03} & $1.70$ & \num{1.89e-03} & $1.49$ \\
\midrule
\multirow{4}{*}{Navier--Stokes} & \multirow{4}{*}{Diffusive} & $\rho$ & Double & 512 & \num{8.56e-07} & $2.03$ & \num{4.28e-07} & $2.02$ \\*
 &  &  & Float & 512 & \num{8.34e-07} & $2.12$ & \num{2.83e-07} & $2.22$ \\*
 &  & $\rho\mathbf{u}$ & Double & 512 & \num{3.73e-04} & $2.00$ & \num{3.72e-04} & $2.00$ \\*
 &  &  & Float & 512 & \num{2.45e-04} & $2.19$ & \num{2.51e-04} & $2.18$ \\
\midrule
\multirow{8}{*}{Resistive MHD} & \multirow{8}{*}{Acoustic} & $\rho$ & Double & 512 & \num{1.51e-04} & $1.84$ & \num{7.41e-05} & $1.87$ \\*
 &  &  & Float & 512 & \num{1.49e-04} & $1.85$ & \num{8.41e-05} & $1.81$ \\*
 &  & $\rho\mathbf{u}$ & Double & 512 & \num{2.34e-04} & $1.86$ & \num{1.36e-04} & $1.90$ \\*
 &  &  & Float & 512 & \num{2.15e-04} & $1.91$ & \num{1.22e-04} & $1.95$ \\*
 &  & $E$ & Double & 512 & \num{2.77e-04} & $1.89$ & \num{1.81e-04} & $1.91$ \\*
 &  &  & Float & 512 & \num{2.32e-04} & $1.97$ & \num{1.48e-04} & $2.00$ \\*
 &  & $\mathbf{B}$ & Double & 512 & \num{1.83e-04} & $2.02$ & \num{1.26e-04} & $2.05$ \\*
 &  &  & Float & 512 & \num{1.71e-04} & $2.05$ & \num{1.24e-04} & $2.06$ \\
\midrule
\multirow{6}{*}{HCNSF (3D)} & \multirow{6}{*}{Acoustic} & $\rho$ & Double & 256 & \num{2.60e-04} & $1.92$ & \num{9.87e-05} & $1.94$ \\*
 &  &  & Float & 256 & \num{5.32e-04} & $1.86$ & \num{2.07e-04} & $1.88$ \\*
 &  & $\rho\mathbf{u}$ & Double & 256 & \num{8.87e-05} & $2.00$ & \num{4.54e-05} & $1.98$ \\*
 &  &  & Float & 256 & \num{1.93e-04} & $1.94$ & \num{9.82e-05} & $1.93$ \\*
 &  & $E$ & Double & 256 & \num{1.18e-04} & $1.88$ & \num{5.50e-05} & $1.91$ \\*
 &  &  & Float & 256 & \num{2.30e-04} & $1.93$ & \num{1.08e-04} & $1.92$ \\
\bottomrule
\end{tabular}
\end{table}

\begin{table}[tbp]
\centering
\footnotesize
\caption{Reconstruction against the gradient-tracking scheme of~\cite{pde2lbm} at matched scaling and
resolution, in double precision. Relative $L_2$ error of each conserved field at $N = 512$, and at
$N = 256$ for the three-dimensional HCNSF system, and the ratio of the gradient-tracking error to the reconstruction
error. A ratio above one is a lower reconstruction error.}
\label{tab:mms_compare}
\begin{tabular}{l l c c c}
\toprule
System & Field & Reconstruction & Tracking~\cite{pde2lbm} & Ratio \\
\midrule
Scalar ADR      & $c$              & \num{1.03e-04} & \num{1.27e-04} & $1.23$ \\
Allen--Cahn     & $\phi$           & \num{5.89e-04} & \num{7.86e-04} & $1.34$ \\
Navier--Stokes  & $\rho$           & \num{4.28e-07} & \num{6.07e-07} & $1.42$ \\
                & $\rho\mathbf{u}$ & \num{3.72e-04} & \num{8.54e-04} & $2.30$ \\
Resistive MHD   & $\rho$           & \num{7.41e-05} & \num{1.63e-04} & $2.20$ \\
                & $\rho\mathbf{u}$ & \num{1.36e-04} & \num{2.45e-04} & $1.80$ \\
                & $E$              & \num{1.81e-04} & \num{2.12e-04} & $1.17$ \\
                & $\mathbf{B}$     & \num{1.26e-04} & \num{1.63e-04} & $1.29$ \\
\midrule
HCNSF ($N = 256$) & $\rho$          & \num{9.87e-05}  & \num{8.45e-05}  & $0.86$ \\
                & $\rho\mathbf{u}$ & \num{4.54e-05}  & \num{5.58e-05}  & $1.23$ \\
                & $E$              & \num{5.50e-05}  & \num{7.43e-05}  & $1.35$ \\
\bottomrule
\end{tabular}
\end{table}

In double precision every conserved field of every system converges at second order (Table~\ref{tab:mms_eoc}).
Against the gradient-tracking scheme of the approach at equal resolution the reconstruction is competitive and on average slightly more accurate, its relative $L_2$ error lower by about a factor $1.4$ across the conserved fields, behind only on the HCNSF density (Table~\ref{tab:mms_compare}).
It reaches that accuracy carrying only the conserved fields, three against the seven of the gradient-tracking scheme on Navier--Stokes and six against eighteen on magnetohydrodynamics.

The single-precision runs inherit the nondimensionalization and reference-state shifting of~\cite{pde2lbm}, which express each field in physical (SI) units and subtract its reference value before collision.
In single precision the resistive magnetohydrodynamics and Navier--Stokes systems keep second-order convergence at close to the double-precision error, and HCNSF at about twice it.
The two lightest diffusively refined scalars, advection--diffusion--reaction and Allen--Cahn, fall below second order in single precision on the finest grids (Table~\ref{tab:mms_eoc}), while double precision holds second order throughout.

\section{Performance}\label{sec:performance}

A lattice Boltzmann kernel is commonly bound by memory bandwidth rather than floating-point throughput~\cite{Krueger2016, Tolke2010, Januszewski2014, Godenschwager2013, Bauer2021walberla, Bauer2021lbmpy}, so the transported field count sets its throughput.
By reconstructing the constitutive gradients rather than transporting them, the scheme carries only the conserved fields, whereas gradient tracking appends auxiliary fields for the tracked gradients.
We benchmark the reconstruction's bare collide-and-stream operators, without statistics or manufactured source terms on an NVIDIA RTX A5000 (\texttt{sm\_86}, measured \qty{680}{\giga\byte\per\second} peak bandwidth, BabelStream Triad~\cite{Deakin2018}).
Results are compared against the published gradient-tracking figures of~\cite{pde2lbm}, obtained on the same hardware.
The kernels are targeted at the hardware-abstracted framework of OpenLB~\cite{Krause2021a, Kummerlaender2023}, which compiles a single-source implementation to diverse CPU and GPU targets~\cite{Kummerlaender2026c}.
Table~\ref{tab:roofline} reports the throughput in \emph{Million Lattice Updates per Second} (MLUP/s), the utilized bandwidth relative to peak, and the per-thread register count, local-memory spill and register-limited occupancy of the bulk collide kernel.
The byte traffic per cell is $2\,q\,n_{\mathrm{fields}}\,\mathrm{sizeof}(T)$, with $q$ the number of discrete velocities of the stencil, counting the read and the write of the in-place populations.

\begin{table}[tbp]
\centering
\scriptsize
\setlength{\tabcolsep}{3pt}
\renewcommand{\arraystretch}{1.15}
\caption{Throughput of the gradient reconstruction against the gradient-tracking scheme
of~\cite{pde2lbm}. The reconstruction is timed on the bare operator (statistics and manufactured source
removed) on an NVIDIA RTX A5000 (\texttt{sm\_86}, measured \qty{680}{\giga\byte\per\second} peak
bandwidth, BabelStream Triad~\cite{Deakin2018}). Byte/cell is
$2\,q\,n_{\mathrm{fields}}\,\mathrm{sizeof}(T)$, with $q$ the number of discrete velocities, counting the
read and the write of the in-place populations. FLOP/cell is the arithmetic-operation count of the emitted collide, loads and
stores excluded. BW is the utilized bandwidth relative to this measured peak. Reg, Spill and Occ are the bulk collide
kernel's per-thread register count, local-memory spill (bytes) and register-limited occupancy, from
\texttt{ptxas}. The gradient-tracking figures are the published values of~\cite{pde2lbm}. Their FLOP/cell can differ between precisions because~\cite{pde2lbm} reports the faster of two emitted CSE variants at each precision. Speedup is the reconstruction MLUP/s over the
gradient tracking at the same system and precision.}
\label{tab:roofline}
\resizebox{\linewidth}{!}{%
\begin{tabular}{l l l c c c c c c c c c}
\toprule
System & Prec. & Scheme & Fields & Byte/cell & FLOP/cell & MLUP/s & BW & Reg & Spill & Occ.\ (\%) & Speedup \\
\midrule
\multirow{4}{*}{Scalar ADR} & \multirow{2}{*}{Double} & Tracking & 3 & 240 & 141 & 1673 & 59\% & 58 & -- & 67 &  \\*
 &  & Recon. & 1 & 80 & 53 & 3843 & 45\% & 40 & -- & 100 & 2.30$\times$ \\*
 & \multirow{2}{*}{Single} & Tracking & 3 & 120 & 253 & 5405 & 95\% & 58 & -- & 67 &  \\*
 &  & Recon. & 1 & 40 & 53 & 13340 & 78\% & 40 & -- & 100 & 2.47$\times$ \\
\midrule
\multirow{4}{*}{Allen--Cahn} & \multirow{2}{*}{Double} & Tracking & 3 & 240 & 124 & 1918 & 68\% & 64 & -- & 67 &  \\*
 &  & Recon. & 1 & 80 & 50 & 4182 & 49\% & 40 & -- & 100 & 2.18$\times$ \\*
 & \multirow{2}{*}{Single} & Tracking & 3 & 120 & 245 & 5487 & 97\% & 64 & -- & 67 &  \\*
 &  & Recon. & 1 & 40 & 50 & 13346 & 79\% & 40 & -- & 100 & 2.43$\times$ \\
\midrule
\multirow{4}{*}{Navier--Stokes} & \multirow{2}{*}{Double} & Tracking & 7 & 560 & 387 & 662 & 55\% & 126 & -- & 33 &  \\*
 &  & Recon. & 3 & 240 & 150 & 865 & 31\% & 86 & -- & 48 & 1.31$\times$ \\*
 & \multirow{2}{*}{Single} & Tracking & 7 & 280 & 387 & 2328 & 96\% & 126 & -- & 33 &  \\*
 &  & Recon. & 3 & 120 & 150 & 5335 & 94\% & 86 & -- & 48 & 2.29$\times$ \\
\midrule
\multirow{4}{*}{Resistive MHD} & \multirow{2}{*}{Double} & Tracking & 18 & 1440 & 2141 & 168 & 36\% & 255 & 16 & 17 &  \\*
 &  & Recon. & 6 & 480 & 750 & 324 & 23\% & 208 & -- & 19 & 1.93$\times$ \\*
 & \multirow{2}{*}{Single} & Tracking & 18 & 720 & 2141 & 736 & 78\% & 255 & 32 & 17 &  \\*
 &  & Recon. & 6 & 240 & 750 & 2746 & 97\% & 208 & -- & 19 & 3.73$\times$ \\
\midrule
\multirow{4}{*}{HCNSF (3D)} & \multirow{2}{*}{Double} & Tracking & 14 & 1568 & 2234 & 163 & 38\% & 255 & 104 & 17 &  \\*
 &  & Recon. & 5 & 560 & 680 & 248 & 20\% & 254 & -- & 17 & 1.52$\times$ \\*
 & \multirow{2}{*}{Single} & Tracking & 14 & 784 & 2234 & 546 & 63\% & 255 & 104 & 17 &  \\*
 &  & Recon. & 5 & 280 & 680 & 2033 & 84\% & 236 & -- & 17 & 3.72$\times$ \\
\bottomrule
\end{tabular}%
}
\end{table}

\begin{figure}[tbp]
\centering
\includegraphics[width=0.82\linewidth]{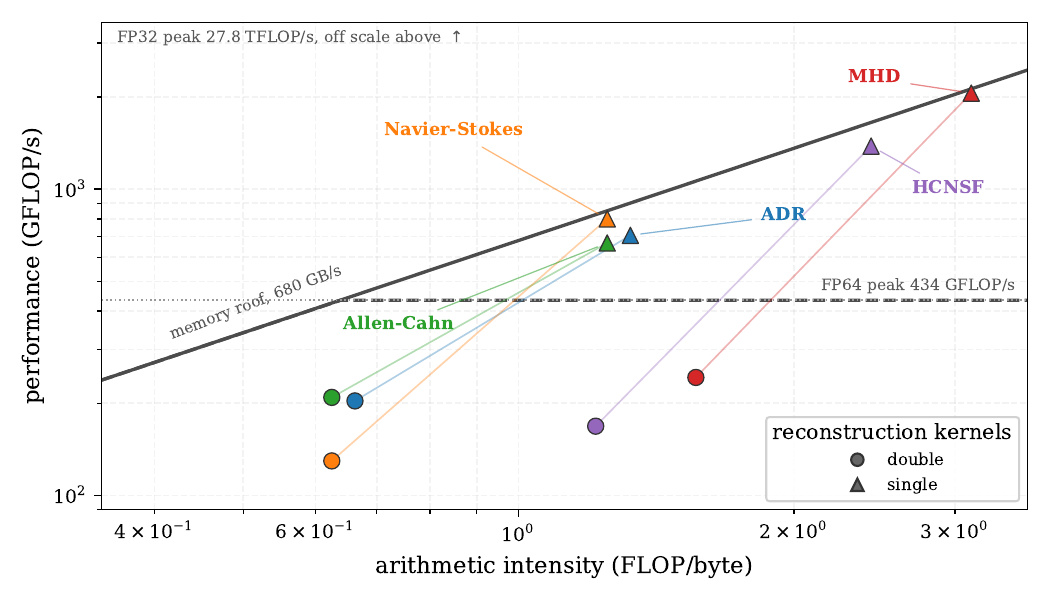}
\caption{Roofline of the generated kernels on the NVIDIA RTX A5000 (memory roof
\qty{680}{\giga\byte\per\second} and FP64 compute ceiling, with the FP32 ceiling an order of magnitude
above the fastest kernel and off the top of the plot). Each kernel is placed at its
arithmetic intensity and attained performance, a circle for double and a triangle for single precision,
joined per system. In single precision the reconstruction kernels lie on or near the memory roof.}
\label{fig:roofline}
\end{figure}

The reconstruction is faster for every target equation in both precisions (Table~\ref{tab:roofline}), the improvement scaling with the reduction in transported fields.
Due to holding fewer populations it uses fewer registers which increases occupancy and avoids the local-memory spills that otherwise occur for the larger gradient-tracking kernels.
It also does less arithmetic per cell since dropping fields eliminates more relaxation work than the moment, operator and closure evaluations add.
Byte traffic falls by up to a factor of three, and in the memory-bound single-precision regime this reduction sets the speedup (Figure~\ref{fig:roofline}).

\section{Conclusion}
We added a gradient reconstruction to the automatic lattice Boltzmann derivation of~\cite{pde2lbm}, invoked by a single declaration and generic across constitutive closures of gradient form.
The gradients a closure requires are recovered from the non-equilibrium moment inside the collision, without any auxiliary state.
The non-equilibrium moment is a linear image of that gradient through the diffusive flux, and the reconstruction is its algebraic inverse.

Across the gradient-tracking systems of the approach, advection--diffusion--reaction, Allen--Cahn, Navier--Stokes, resistive magnetohydrodynamics and the three-dimensional homogenized compressible Navier--Stokes--Fourier system, the scheme converges at second order in double precision, competitive with the gradient-tracking scheme at equal resolution while carrying only the conserved fields.
In single precision the reference-shifted formulation of~\cite{pde2lbm} keeps it at or near second order (Section~\ref{sec:convergence}).
On an NVIDIA RTX A5000 it is up to $3.7$ times faster in single precision, the memory-bound kernels reaching up to $97\%$ of peak bandwidth.
Removing the auxiliary gradient fields also removes the register spills of the magnetohydrodynamics and HCNSF gradient-tracking kernels.

\section*{Code Availability}

The gradient reconstruction is specified by Section~\ref{sec:methodology}, and the manufactured solutions, lattices and refinement laws are those of~\cite{pde2lbm}.
The \emph{PDE2LBM} compiler is available upon reasonable request.

\section*{Acknowledgments}

Google Gemini and Anthropic Claude assisted in drafting and revising the manuscript.
All content was reviewed, verified, and approved by the authors, who take full responsibility for it.

% Flush any pending full-page floats before the bibliography so they cannot
% drift past it, while letting them place naturally within the text (no forced gaps).
\FloatBarrier
\bibliographystyle{siamplain}
\bibliography{literature}

\end{document}